\documentclass[apj,onecolumn]{emulateapj}

\newcommand{\kev}{keV}
\newcommand{\fe}{Fe~K$\alpha$}
\newcommand{\etal}{et al.}

\newcommand{\fouru}{4U~1820--30}

\shorttitle{Suddenly Heated Accretion Disks around Neutron Stars}
\shortauthors{Ballantyne \& Everett}

\usepackage{times}
\usepackage{graphicx}

\begin{document}

\title{On the Dynamics of Suddenly Heated Accretion Disks around
  Neutron Stars}


\author{D. R. Ballantyne and J. E. Everett}
\affil{Canadian Institute for Theoretical Astrophysics,
McLennan Labs, 60 St. George Street, Toronto, Ontario, Canada M5S 3H8}
\email{ballantyne, everett@cita.utoronto.ca} 

\begin{abstract}
Type~I X-ray bursts and superbursts on neutron stars release sudden
and intense radiation fields into their surroundings. Here, we
consider the possible effects of these powerful explosions on the
structure of the accretion disk. The goal is to account for the
apparent evolution of the innermost regions of the accretion disk
around \fouru\ during a superburst. Three different processes are
considered in detail: radiatively or thermally driven outflows, inflow
due to Poynting-Robertson drag, and a structural change to the disk by
X-ray heating. Radiatively driven winds with large column densities
can be launched from the inner disk, but only for $L/L_{\mathrm{Edd}}
\ga 1$, which is expected to be obtained only at the onset of the
burst. Furthermore, the predicted mass outflow rate is less than the
accretion rate in \fouru. Estimates of the Poynting-Robertson or
radiative drag timescale shows that it is a very efficient means of
removing angular momentum from the gas. However, the analytical
results are likely only applicable at the innermost edge of the
disk. X-ray heating gives a change in the disk scale height that is
correlated with the blackbody temperature, as seen in the evolution
during the \fouru\ superburst. If this change in the scale height can
alter the surface density, then the viscous time (with $\alpha \sim
0.03$--$0.2$) is the closest match to the \fouru\ results. We expect,
however, that all three processes are likely ongoing when an accretion
disk is subject to a sudden heating event. Ultimately, a numerical
simulation of a disk around a bursting neutron star is required to
determine the exact response of the disk. Magnetic truncation of the
accretion flow is also considered and applied to the \fouru\ X-ray reflection
results.
\end{abstract}


\section{Introduction}
\label{sect:intro}
Accretion disks around neutron stars are often subject to sudden
intense radiation fields due to Type I X-ray bursts
\citep*{lvt93,lvt95,bil98,sb03,cum04}, or, on rare occasions, superbursts
\citep{cor00,cb01,sb02,kuul04}. These events can reach the Eddington
luminosity, and thus deposit a large amount of energy in the accretion
disk over a short amount of time (tens of seconds for Type I bursts;
hours for superbursts). The accretion disk structure may be altered
by this sudden injection of energy, as suggested by the evolution of
X-ray reflection features observed from the low-mass X-ray binary
(LMXB) \fouru\ during a superburst \citep{bs04}. In this source,
modeling of \fe\ line and edge features produced by the disk indicated
that the reflecting part of the disk moved outward from $\sim
20$~$r_g$ ($r_g=GM/c^2$, where $M$ is the mass of the neutron star) to
$\sim 100$~$r_g$ in $\sim 1000$~s, before eventually returning to
$\sim 10$~$r_g$. This behavior was puzzling because the change in the
reflecting radius was not correlated with the observed X-ray flux, but
with the hardness of the blackbody spectrum. \citet{bs04}
therefore proposed a model in which the surface density of the inner
accretion disk was altered by heating from the incident X-rays;
however, alternative models have not yet been carefully considered.

In this paper, we examine in detail a number of different physical
processes which may alter the structure of accretion disks during a
sudden heating event. The physics are treated as generally as
possible, but quantitative comparisons are made with the \fouru\
results in order to test the individual models as possible
explanations for the inferred behavior.

We begin in the next section by reviewing the observed properties of
\fouru.  The paper is then broken into three sections based on the
different possibilities for the accretion disk response. In
\S~\ref{sect:outflows} we consider outflow mechanisms, such as
thermally and radiatively driven winds. \S~\ref{sect:inflow} contains a
study of inflow onto the central object via radiative drag. Changes to
the disk structure that do not involve inflow or outflow are the
subject of \S~\ref{sect:heating}. The results are gathered and
discussed in \S~\ref{sect:discuss}. Finally, conclusions are drawn in
\S~\ref{sect:concl}.

\section{Overview of \fouru}
\label{sect:overview}
\fouru\ is one of the most compact LMXBs known with an orbital period
of only 11.4 minutes \citep*{spw87}. Evolutionary models of the system
\citep*{ver87,bg87,rap87,rpr00} all predict that the accreting material
is primarily helium. Type I X-ray bursts from \fouru\ have been
observed for nearly 30 years \citep*{gri76,hab87,kuul03}, and have
recurrence times between 2 and 4 hours. \citet{cum03} presents further
details and models of the Type I X-ray bursts from \fouru.

Although the system resides within the metal rich globular cluster
NGC~6624, the distance to \fouru\ is still not precisely known. The
difficulty arises because of uncertainties in the interstellar
extinction along the line-of-sight, and in the metallicity relation
used to correct the absolute magnitude of horizontal branch
stars. Thus, using the same technique, \citet{kuul03} derived the
distance to \fouru\ of 7.6$\pm$0.4~kpc, in agreement with an earlier
estimate \citep{rml93}, while \citet*{vlv86} found it
to be 6.4$\pm0.6$~kpc. Most recently, \citet{st04} modeled the light-curve and
spectrum of a photospheric radius expansion (PRE) burst from \fouru\
and determined a distance of $\sim 5.8$~kpc.

\begin{figure}
\centerline{
\includegraphics[angle=-90,width=0.9\textwidth]{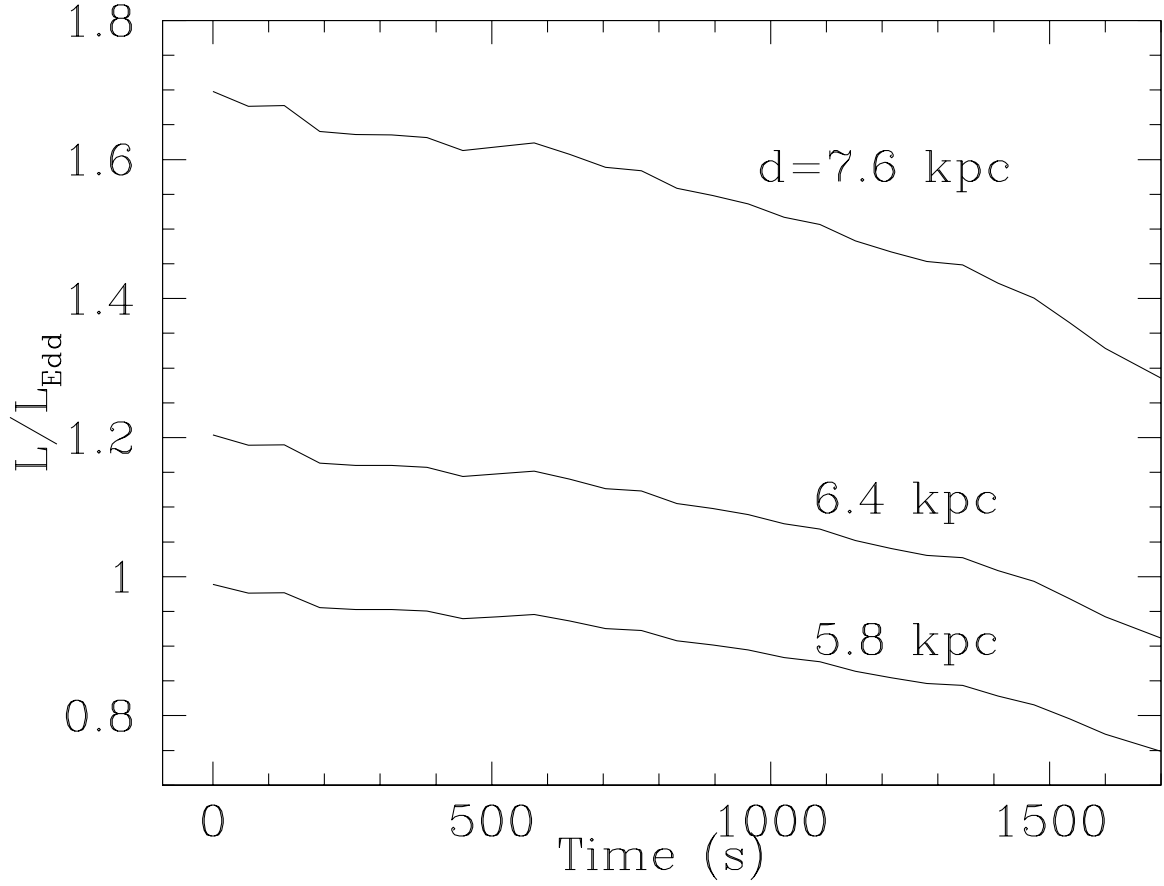}
}
\caption{Time evolution of $L/L_{\mathrm{Edd}}$ during the first
  $\sim$1700~s of the the \fouru\ superburst. The three curves show
  the results for different distance estimates to the
  system. $L_{\mathrm{Edd}}$ is taken to be $2.7\times
  10^{38}$~ergs~s$^{-1}$, assuming a canonical neutron star
  ($M=1.4$~M$_{\odot}$, $R=10$~km) with a helium atmosphere. The
  observed luminosity was calculated between 0.1--40~\kev\ using the
  best-fitting spectral models of \citet{bs04} and assuming isotropic
  radiation. }
\label{fig:loverledd}
\end{figure}

The Eddington luminosity at the photosphere of a neutron star with a
helium atmosphere as determined by a distant observer is
\begin{equation}
L_{\mathrm{Edd}}=2.5\times
10^{38}\left ( {M \over \mathrm{M_{\odot}}} \right ) \left [
  1-0.295 \left ( {M \over \mathrm{M}_{\odot}} \right) \left(
  {R_{\ast} \over 10\ \mathrm{km}} \right )^{-1} \right]^{1/2}~\mathrm{ergs~s}^{-1},
\label{eq:ledd}
\end{equation}
where $R_{\ast}$ is the radius of the photosphere
\citep{lvt93}. Employing the standard assumptions (which will be made
throughout this paper) of a 1.4~M$_{\odot}$, 10~km radius neutron
star, $L_{\mathrm{Edd}}=2.7\times 10^{38}$~ergs~s$^{-1}$. In
Figure~\ref{fig:loverledd} we plot the time evolution of
$L/L_{\mathrm{Edd}}$ over the first 1600~s of the \fouru\
superburst. The observed luminosity was calculated by measuring the
0.1--40~\kev\ flux predicted from the best fitting spectral models of
\citet{bs04}, and assuming isotropic radiation. Separate curves are
plotted for the different distance estimates described above.  Since
large super-Eddington luminosities are not expected for X-ray bursts
\citep[e.g.,][]{jm87}, this figure seems to indicate that the 5.8~kpc
distance may be the most reasonable. However, by using
equation~\ref{eq:ledd} we find that increasing the mass and radius of
the neutron star to, for example, 1.7~M$_{\odot}$ and 12~km, would
make the 6.4~kpc distance consistent with
$L_{\mathrm{Edd}}$. The 7.6~kpc distance is more challenging,
but a much larger mass and radius (for example, $M$=2.3~M$_{\odot}$
and $R$=18~km) can allow the Eddington luminosity to approach the
observed value. Such a large mass is
implied by the highest frequency of the kHz QPO discovered from
\fouru\ if it is assumed to correspond to the orbital frequency at the
innermost stable circular orbit \citep{zha98}. Finally, it is worth
noting that since the superburst had a PRE phase to it, the emitting
photosphere may be larger than the true stellar radius, resulting in
$L_{\mathrm{Edd}}$ being underestimated, and therefore
overestimating $L/L_{\mathrm{Edd}}$. Because of these
uncertainties, we will, if necessary, use all three values of the
distance in the following sections, and continue to assume
$L_{\mathrm{Edd}}=2.7\times 10^{38}$~ergs~s$^{-1}$.

X-ray bursts are seen from \fouru\ when the persistent X-ray flux is
$\la 4\times 10^{-9}$~ergs~cm$^{-2}$~s$^{-1}$, which is the low end of
the observed range. This emission is powered by accretion onto the
neutron star, so the flux can be converted to an accretion rate, $\dot
M \approx 1.1\times 10^{17} (d/6.4\ \mathrm{kpc})^2$~g~s$^{-1}\ =
1.7\times 10^{-9} (d/6.4\ \mathrm{kpc})^2$~M$_{\odot}$~yr$^{-1}$ \citep{sb02},
consistent with the prediction of gravitational radiation driven
mass-loss \citep{spw87}. Using the above value of
$L_{\mathrm{Edd}}$ and the observed persistent luminosity, we
obtain an Eddington fraction of $\dot M/\dot M_{\mathrm{Edd}} = 0.07
(d/6.4\ \mathrm{kpc})^2$, where $\dot
M_{\mathrm{Edd}}=L_{\mathrm{Edd}}/\eta c^2$ is the Eddington mass
accretion rate ($\eta$ is the accretion efficiency).

\section{Outflows}
\label{sect:outflows}

\subsection{Radiatively driven winds}
\label{sub:radiative}
Type I X-ray bursts and superbursts may sometimes reach the Eddington
luminosity $L_{\mathrm{Edd}}$, as witnessed in the radiation expansion
phase, where the photosphere of the burning material can be lifted off
of the neutron star \citep*[e.g.,][]{lvb84,sb03}. At these luminosities the
momentum in the radiation field may also affect the inner regions of the
accretion disk, possibly even blowing it away from the central
object. Below, we check if a radiatively driven wind can
explain the behavior of \fouru\ during the superburst.

\subsubsection{Estimates from Accretion Disk Theory}
\label{subsub:estimates}
Before diving into the models of radiative winds, it is worth
calculating some simple estimates of outflow rates and 
energetics from basic \citet{ss73} accretion disk theory to compare
against the computations.

The X-ray modeling showed that the reflecting region moved from $\sim
20$~$r_g$ to $\sim 100$~$r_g$ in 500--1000~s \citep{bs04}. It is
unknown what occurred to the material within 20~$r_g$ to prevent it
from reflecting (this is speculated upon in \S~\ref{sect:discuss}), so
we will only consider the evolution of the disk between 20 and
100~$r_g$.  The accretion rate onto the neutron star is low enough
(for all three values of the distance) that almost all of the
radiation pressure dominated region of the disk will be within
20~$r_g$ \citep{ss73}, thus we will work only with the equations for a
gas supported disk with electron scattering opacity. In this domain,
the surface density of the disk can be written as
\begin{equation}
\Sigma=2.5\times10^{13} (\mu m_{\mathrm{p}})^{4/5} (0.2(1+X))^{-1/5}
\alpha^{-4/5} \left ( {M \over \mathrm{M_{\odot}}} \right )^{-2/5} \dot M^{3/5}
\left ( {R \over r_g} \right )^{-3/5} J(R)^{3/5}\ \mathrm{g\
    cm^{-2}},
\label{eq:sigma}
\end{equation}
where $R$ is the radius along the disk, $\mu$ is the mean molecular
weight of the accreting gas, $m_{\mathrm{p}}$ is the proton mass, $X$
is the mass fraction of hydrogen, $\alpha$ is the Shakura-Sunyaev
viscosity parameter, and $J(R)=(1-(6 r_g/R)^{1/2})$
\citep[e.g.,][]{sz94}. For \fouru, $\dot
M \approx 1.1\times 10^{17} (d/6.4\ \mathrm{kpc})^2$~g~s$^{-1}$, $X
\approx 0$ and $\mu \approx 1.3$ for a fully ionized gas with a cosmic
abundance of heavy metals. This results in 
\begin{equation}
\Sigma=6\times10^4 \alpha^{-4/5} \left ( {M \over 1.4\ 
  \mathrm{M_{\odot}}} \right )^{-2/5} \left ( {\dot M \over 1.1\times
  10^{17}\ \mathrm{g}\ \mathrm{s}^{-1}} \right )^{3/5} \left ( {d
  \over 6.4\ \mathrm{kpc} } \right )^{6/5} \left ( {R \over r_g}
  \right )^{-3/5} J(R)^{3/5}\ \mathrm{g\ cm^{-2}}. 
\label{eq:surfdens}
\end{equation}
We then integrate eq.~\ref{eq:surfdens}
between 20~$r_g$ and 100~$r_g$ to estimate the total mass in the disk
that needs to be removed by an outflow, $m = 3.2\times 10^{19} (d/6.4\
\mathrm{kpc})^{6/5}$~g, where we have assumed $\alpha =0.1$. If this
mass is removed in $\sim 500$~s, then we obtain a mean mass outflow
rate of $\bar{\dot m}=6.4\times 10^{16} (d/6.4\
\mathrm{kpc})^{6/5}$~g~s$^{-1}$. This is the same order of magnitude
as the accretion rate onto the neutron star, which implies that an
even higher outflow rate is needed in order to remove the inner disk.

If this material is initially being lifted off an area $A$ of the disk
at the sound speed $c_s$ then by mass continuity, $\bar{\dot m}= \rho c_s
A=\bar{n} m_{\mathrm{He}} c_s A$ where $\bar{n}$ is the mean number
density of helium at the disk surface. Writing the sound speed as $c_s
= \sqrt{kT/\mu m_\mathrm{p}}$, and taking $kT = 2.6$~keV, the
temperature of the blackbody near the beginning of the superburst
\citep{sb02}, gives $c_s \approx 435$~km~s$^{-1}$. If the launching
area is an annulus between 20 and 100~$r_g$, then $\bar n=1.7\times
10^{17} (d/6.4\ \mathrm{kpc})^{6/5}$~cm$^{-3}$. This is much lower than
the number density predicted by \citet{ss73}, but may be reasonable
for the surface layers where the density must fall off rapidly.

Equation~\ref{eq:surfdens} can also be used to calculate the
gravitational potential energy of the disk material between 20 and
100~$r_g$, $5.2\times 10^{38} (d/6.4\
\mathrm{kpc})^{16/5}$~ergs. The energy released during the
first $\sim 1500$~s of the \fouru\ superburst is $10^{41-42}$~ergs,
assuming isotropic radiation. Therefore, there seems to be enough energy in
the radiation field to completely remove the reflecting region of the
accretion disk in a radiatively driven wind. The viability of such a
wind is considered in more detail below.

\subsubsection{Numerical Models}
\label{subsub:models}
A numerical wind model can be used to calculate the radiative
acceleration, velocity, and escape time for outflowing gas.  For this
paper, we define a model of a continuum-driven wind, launched from the
accretion disk, and moving along radial streamlines (radial because
the wind is driven by momentum taken from radially-streaming photons).
This outflow structure therefore fills an open bi-cone in which gas
flows radially outward, nearly along the surface of the accretion disk
(for simplicity the wind is set here to flow at a constant $5^{\circ}$
above the disk).

The wind's acceleration is computed by taking the local velocity,
density, ionization structure, and transmitted continuum into account.
The outflow is launched from a user-defined radius with a nearly
optically thick column at the disk (the maximum column that a
radiation driving could support), with the initial outflow speed set
to the sound speed.  The photoionization of this outflowing gas is
then simulated using Cloudy \citep{Ferland02} at various heights in
the wind.  These simulations yield both the transmitted continuum and
the ionization state of the gas from which the radiative acceleration
can be calculated.  Integrating this acceleration then yields the
solution for the velocity as a function of distance along the wind.
We then iterate on this solution to ensure that the density,
photoionization, acceleration, and velocity profiles are all
consistent\footnote{This model follows largely the same method as
presented in \citet{eb04}.  In contrast to that earlier work, however,
the angular momentum of the orbiting gas in these simulations is set
solely by the mass of the central object (since the initial angular
momentum was set by the quiescent luminosity, not the outburst
luminosity modeled here).}.

The initial conditions for the model are chosen to match the observed
and modeled properties of 4U~1820-30 \citep{sb02, bs04}. First, the
gas is illuminated with a 2.6~keV blackbody continuum, matching the
temperature inferred by \citet{sb02} near the peak of the observed
flux. We use gas abundances for cool extreme helium stars as given by
\citet{Pandey01}. The initial number density was set to $10^{17}~{\rm
cm}^{-3}$ at the disk's surface, as implied by the simple calculation
in the previous section\footnote{Our assumed base density is much
greater than the $10^{13}~{\rm cm}^{-3}$ limit beyond which Cloudy's
three-body recombination calculations are defined \citep{Ferland02}.
However, for these luminosities, the ionization level is so high ($U =
n_{\rm ion}/n_{\rm He} \sim 10^{4.5}$, where $n_{\mathrm{ion}}$ is the
density of ionizing photons) that this will not substantially affect
our results.}.  The innermost streamline of the wind is set to a
radius of 20~$r_g$, the initial distance derived from the \fe\
line. As mentioned above, the base of the wind is simulated as being
approximately optically thick with $N_H = 10^{24}~{\rm cm}^{-2}$ to
model the entire wind that would be launched radiatively from the disk
and to set a conservative limit on the velocity of the outflowing gas.
At the specified initial density, this column implies wind launching
radii of 20 to $\sim$70~$r_g$, approximately the same range as the
observed variation in the \fe\ emission radius.  These models are then
run for a range of $L/L_{\rm Edd}$ values.  For these calculations, it
is important to note that the $L_{\rm Edd}$ defined in
\S~\ref{sect:overview} is replaced by the \textit{local} Eddington
luminosity for the accretion disk gas at $r \gtrsim 50 r_g$ where
relativistic effects are not important.  At those distances, we can
effectively disregard the final bracketed expression in
eq.~\ref{eq:ledd}, yielding $L_{\rm Edd} = 3.5 \times
10^{38}$~ergs~s$^{-1}$. Thus, $L/L_{\rm Edd}$ is set to the local
value that the disk and wind would see.

\begin{figure}
\begin{center}
\plotone{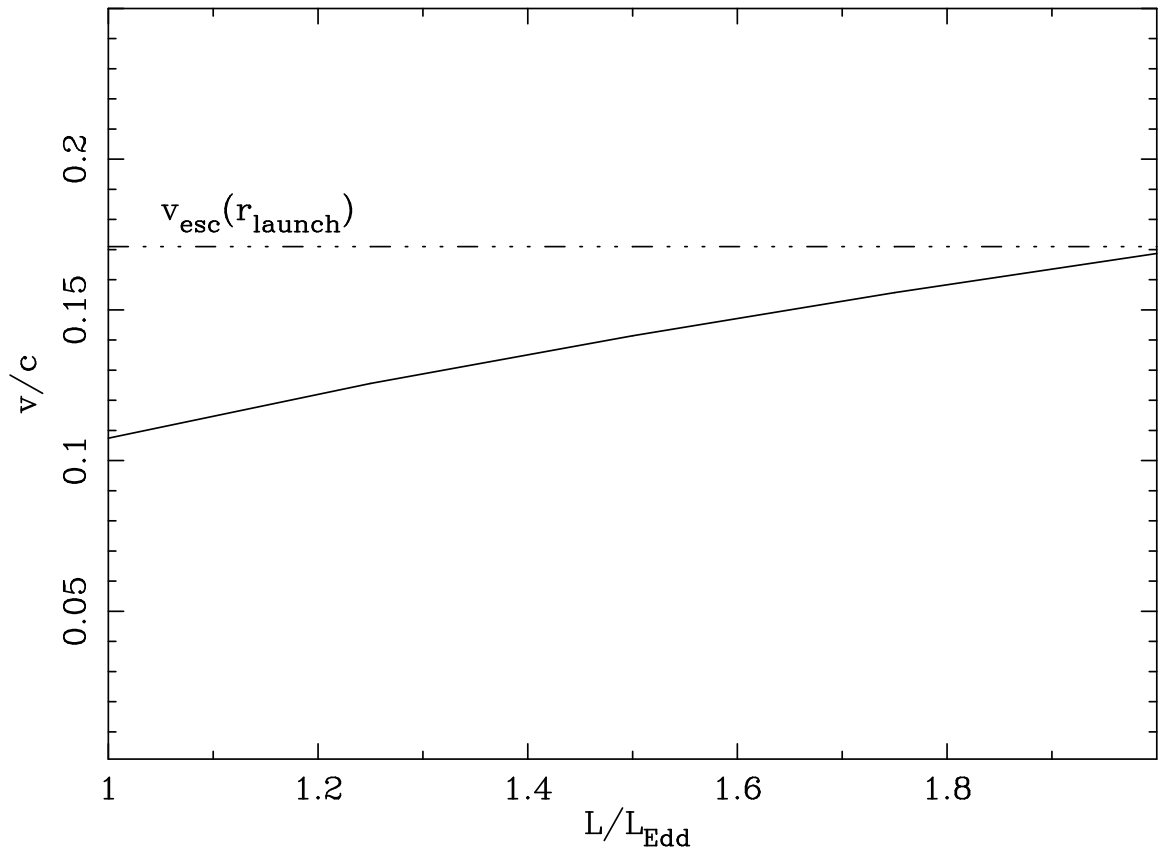}
\caption{Final velocity of a continuum-driven wind with column density
  $N_{\mathrm{H}}=10^{24}$~cm$^{-2}$ as a function of
  the Eddington ratio.  The escape velocity for $r_{\rm launch} \sim
  70$~$r_g = 1.4 \times 10^7 {\rm cm}$ is given with the dot-dashed
  line.\label{vObsPlot}}
\end{center}
\end{figure}

With the above parameters, models are then run for a range of
Eddington ratios to examine the resultant outflow velocities.  These
simulations show that the gas is highly ionized by the superburst,
such that the wind is accelerated only by electron scattering and
bound-free opacity of the wind, justifying \textit{a posteriori} the
approximation that the wind is continuum-driven.
Figure~\ref{vObsPlot} shows the velocity of this wind at
$\sim100$~$r_g$ as a function of the Eddington ratio.



\begin{figure}
\begin{center}
\plotone{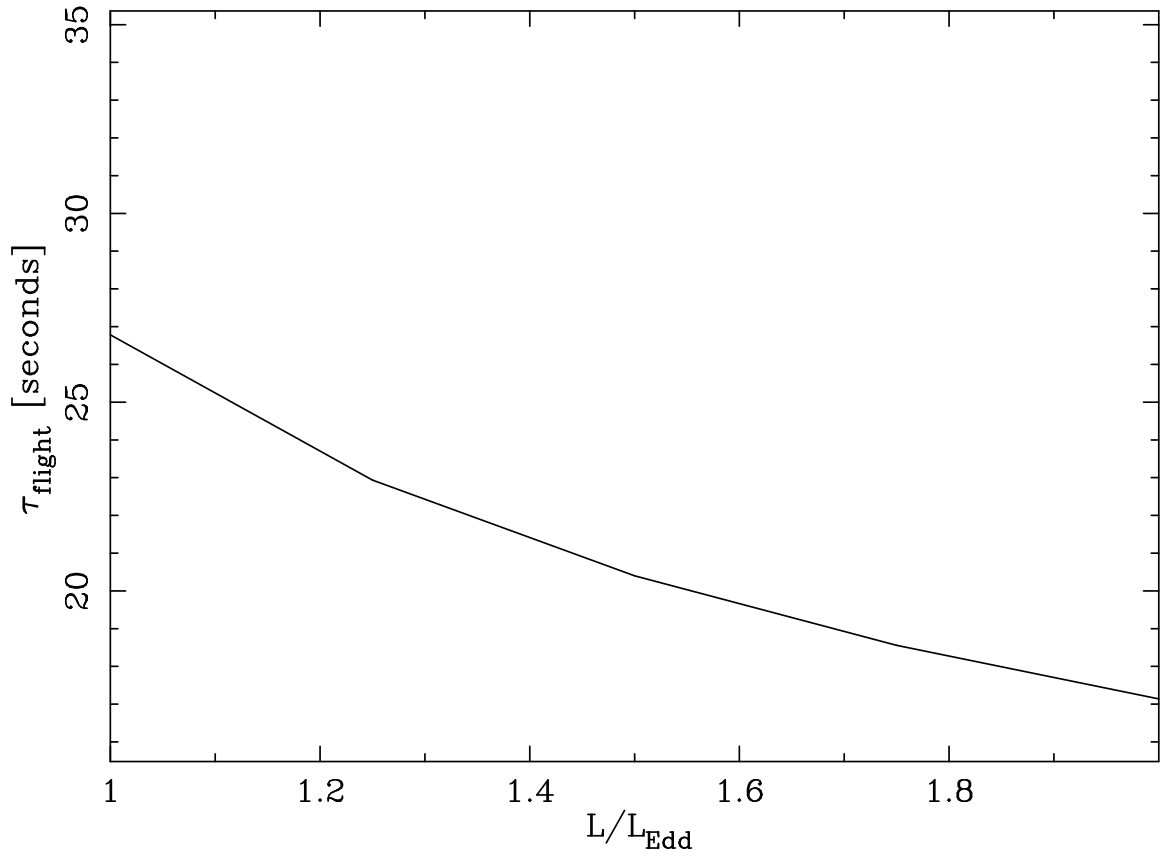}
\caption{Time-of-flight for gas launched at $\sim$20~$r_g$ to reach
  100~$r_g$ from the central source.\label{tauPlot}}
\end{center}
\end{figure}

 The time required for the gas to be launched from 20 to 100~$r_g$ is
therefore calculated by a numerical integral over $dr/v(s)$ where
$v(s)$ is the velocity as a function of distance along the outflow
streamline.  The results are shown in Figure~\ref{tauPlot} for the
model where $N_H = 10^{24}~{\rm cm}^{-2}$ at the base of the wind.
The gas would easily be driven beyond 100~$r_g$ in $\ll \sim
500$--$1000$~s as long as the source has $L/L_{\mathrm{Edd}} \ga 1$.
These results are insensitive to changes in initial conditions, such
as the initial density or blackbody temperature, since variations in
either do not substantially change the ionization state of the gas or
the continuum acceleration.

We now ask whether the mass outflow rate from the disk is enough to
remove a significant amount of mass from the accretion disk.  From the
above model with $n_{\mathrm{He}} = 10^{17}$~cm$^{-3}$, for
$L/L_{\mathrm{Edd}} \sim 1$, $v \sim 0.11c$ and $n_{\rm wind} \sim 1.9
\times 10^{14}~{\rm cm}^{-3}$ at $r = 100$~$r_g = 2.07 \times
10^7~{\rm cm}$.  Calculating $\dot{m}_{\rm out} = 4 \pi r^2 f \rho v$
where $f$ is the covering fraction of the outflow, we find
$\dot{m}_{\rm out} \sim 2.3 \times 10^{16}~f~{\rm g~s}^{-1}$. In
\S~\ref{subsub:estimates}, we estimated from standard accretion theory
that an outflow rate of $\sim 6\times 10^{16}$~g~s$^{-1}$ would be
required to remove the inner part of the disk. The geometry of the
calculated streamlines imply $f \sim 0.1$, resulting in an outflow
rate $\sim 30$ times smaller than the theoretical estimate. Therefore,
these calculations show that it is feasible for a continuum-driven
outflow to remove a significant portion of the inner accretion disk
during a superburst, but only for $L/L_{\mathrm{Edd}} \ga 1$ and it
would not completely empty the inner accretion disk around \fouru\ on
the timescales observed ($\sim 500$~s). The observed mass accretion
rate for this system is $1.1\times 10^{17}~{\rm g~s}^{-1} (d/6.4\
\mathrm{kpc})^2$ (\S~\ref{sect:overview}), implying that the
replenishment rate in the disk is very competitive with the outflow
rate.

\subsection{Thermally driven winds}
\label{sub:thermal}
The gas near the surface of the accretion disk around the neutron star
will be subject to intense X-ray heating by the explosion. If the
temperature of the material increases to the point where the local
sound speed exceeds the escape velocity then a thermal,
pressure-driven wind could form, removing material from the disk. This
mechanism has been proposed as a means to disperse circumstellar gas
around newly forming stars \citep[e.g.,][]{font04}. However, it is
very unlikely that such a wind will form close to a compact object
such as a neutron star, where the escape velocity will be a large
fraction of the speed of light. This can be seen explicitly by
comparing the temperature of the irradiated gas to the virial
temperature $kT_{v} \sim GM\mu m_p/R$. During an outburst by
the central source, the temperature at the accretion disk surface will
be dominated by photoionization, so will maximize at the Compton
temperature. For the case of \fouru, the peak temperature of the
illuminating blackbody was $\sim 3$~keV \citep{sb02}, which will also
be the value of the Compton temperature of the illuminated gas
(\S~\ref{sect:heating}). For gas with $\mu\approx 1.3$ orbiting a
1.4~M$_{\odot}$ neutron star at 20$~r_g$, $kT_v \sim 6\times
10^4$~\kev. Clearly, thermal pressure alone will not cause the heated
gas to flow from the system. Other forces are required to overcome the
strong gravity of the neutron star and cause gas to outflow (\S
\ref{sub:radiative}).

\section{Inflow}
\label{sect:inflow}
X-ray photons produced in the explosion on the surface of the neutron
star can remove angular momentum from the accretion flow by
Poynting-Robertson drag \citep{poy03,rob37,blu74}. This will increase
the local accretion rate and potentially could drain material from the
inner part of the disk inward and onto the star
\citep{wm89,wal92,ml93,ml96}. If this effect operates efficiently, it
may provide another possible explanation for the apparent accretion
disk evolution observed in \fouru.

A simple estimate of the mass affected by radiation drag is when it
interacts with its own rest mass of light, $M_{\mathrm{PR}} \sim
Lt/c^2$, where $t$ is the duration of the event
\citep[e.g.,][]{wm89}. In the case of \fouru, $Lt \sim
10^{41-42}$~erg (\S~\ref{subsub:estimates}) which gives
$M_{\mathrm{PR}} \sim (1.1\times 10^{20-21})f$~g, where $f$ is the
fraction of the luminosity intercepted by the disk. Even if $f \sim
0.1$, this estimate is comparable to the mass in the affected region
of the accretion disk (\S~\ref{subsub:estimates}). We now check the
timescale for angular momentum loss.

For circular orbits with velocity $v_{\phi}$, the rate of change of
angular momentum, $l$, is related to the change of radius $r$ by
\begin{equation}
{1 \over l}{dl \over dt} = {1 \over 2r}{dr \over dt}.
\label{eq:drdt}
\end{equation}
The theory of Poynting-Robertson drag gives the angular momentum loss
rate as a function of the source luminosity $L$ and $r$ \citep[e.g.,][]{blu74}:
\begin{equation}
{dl \over dt} = {- \sigma_{\mathrm{T}} \beta L \zeta f \over 4 \pi c
  r}\Xi,
\label{eq:dldt}
\end{equation}
where $\sigma_{\mathrm{T}}$ is the Thomson cross-section,
$\beta=v_{\phi}/c$, $\zeta$ is the mean number of scattering electrons
per atomic mass units in the gas, $f \approx 5$ is the increase
in the angular momentum loss rate due to relativistic effects \citep{ml96},
and $\Xi$ is a function of photon energy and contains corrections
for Coulomb scattering \citep{blu74}. Substituting eq.~\ref{eq:dldt} into
eq.~\ref{eq:drdt} and integrating gives the time for matter to spiral
from a radius $r_0$ to the stellar surface at $r_{\ast}$:
\begin{equation}
t_{\mathrm{PR}}= {\pi m c^2 (r_0^2 - r_{\ast}^2) \over \sigma_{\mathrm{T}} L \zeta f \Xi},
\label{eq:t1}
\end{equation}
where $m$ is the mass of the scattering particle in the accretion flow.

During the \fouru\ superburst the illuminating spectrum was a
blackbody with a peak temperature of $kT \approx 2.9$~keV
\citep{sb02}. Since the vast majority of the illuminating photons have
energies far above the ionization energy of helium (54.4~eV) $\zeta =
0.5$. \citet{blu74} shows in his Equation (29) a limiting expression for $\Xi$
in the case of a blackbody spectrum with $kT \ll mc^2$. Using this
equation with $kT=2.9$~\kev\ gives $\Xi=0.91$. The radius expansion
phase observed early on in the superburst implies the peak luminosity
was $\sim L_{\mathrm{Edd}}$, which we will take to be $2.7\times
10^{38}$~ergs~s$^{-1}$ (\S~\ref{sect:overview}). Collecting all the
above results and scaling the disk radii to gravitational radii $r_g$
gives a new expression for the time for matter to drain onto the
neutron star due to Poynting-Robertson drag:
\begin{equation}
t_{\mathrm{PR}}=2\times 10^{-6} \left ( {M \over \mathrm{1.4\
M}_{\odot}} \right )^{2}  \left ( {L \over L_{\mathrm{Edd}}} \right )^{-1} \left [ \left ( {r_0 \over r_g}
  \right )^2 - \left ( {r_{\ast} \over r_g} \right )^2 \right]\ s.
\label{eq:t2}
\end{equation}

\begin{figure}
\centerline{
\includegraphics[angle=-90,width=0.9\textwidth]{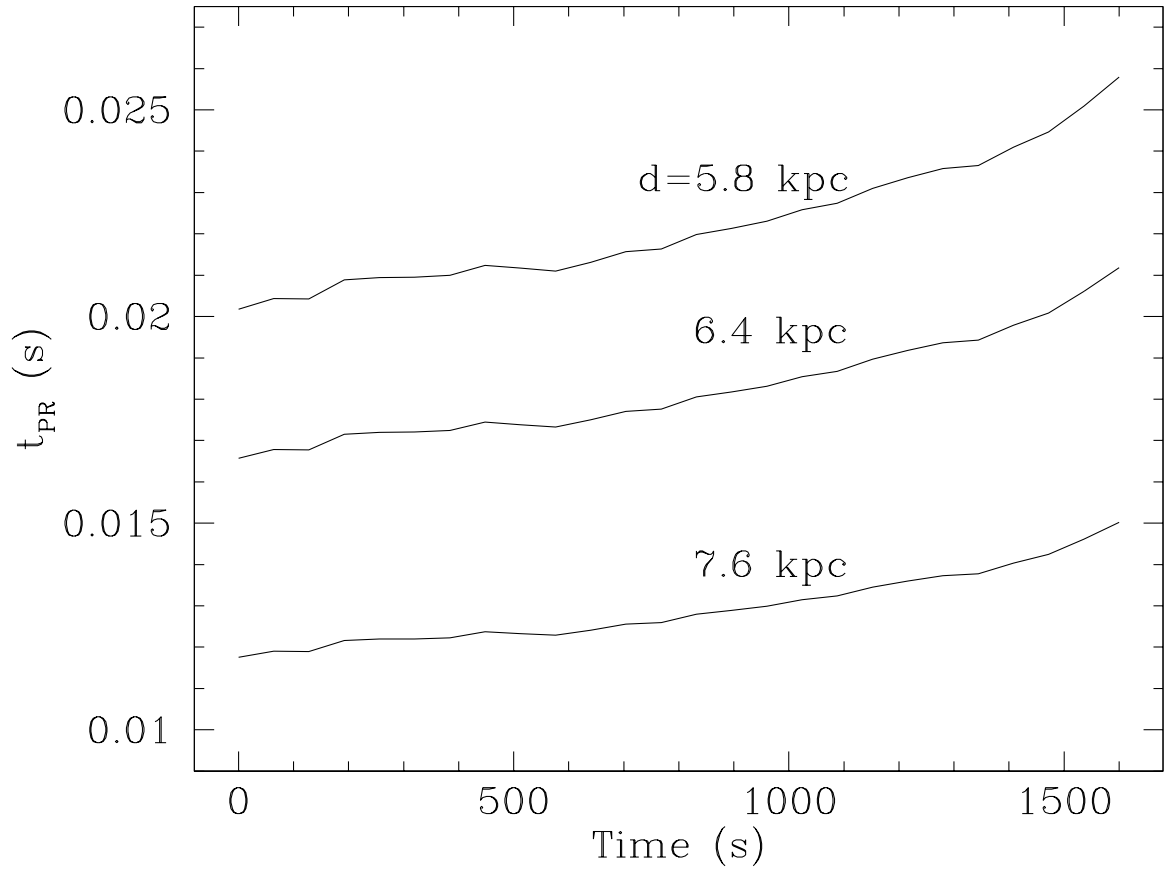}
}
\caption{Timescale for matter to drain onto the neutron star via
  Poynting-Robertson drag during the initial $\sim 1600$~s of the \fouru\
  superburst. As the timescale is directly related to the luminosity,
  three lines are plotted for different estimates of the distance
  to the system. The luminosity was calculated as described in the
  caption to Fig.~\ref{fig:loverledd}. In all three cases,
  $t_{\mathrm{PR}}$ is much less than 1~s.}
\label{fig:poynting}
\end{figure}

Figure~\ref{fig:poynting} plots the predicted time for matter to
spiral down from 100~$r_g$ to the stellar surface at $\sim 5$~$r_g$
during the first $\sim 1600$~s of the \fouru\ superburst.  The values
of $L/ L_{\mathrm{Edd}}$ were taken from Fig.~\ref{fig:loverledd}, and
we show results for the different distance estimates to \fouru.  For
any distance, the Poynting-Robertson timescale is very small compared
to the results of the spectral modeling which showed an evolution in
the reflecting region over 500--1000~s. However, equation~\ref{eq:t2}
ignores processes such as outflows pushed by the radiation field
(\S~\ref{sub:radiative}), viscous coupling of the gas, and radiative
transfer effects in the optically thick disk material which will
randomize the radiation field. All of these will slow down the infall
rate onto the star, but it is unclear if the timescale can be
increased by the four orders of magnitude needed to bring it in line
with the observed behavior. Poynting-Robertson drag also suffers from
the same dependence on the incident flux as the radiative winds in the
previous section. In contrast, the X-ray reflection results show that the disk
response was not dominated by the illuminating flux, but rather by the
spectral shape \citep{bs04}.

Nevertheless, there are indications in the data that the
Poynting-Robertson effect may be ongoing and affecting the disk
structure. \citet{bs04} found that late in the burst decay phase, the
inner radius of the disk (as judged by the reflection features) was
relatively steady between 10 and 20~$r_g$, much farther out than the
inner-most stable circular orbit of $\sim 6$~$r_g$. At this point, the
illuminating spectrum was cool and the observed flux was about 10
times lower than at the peak. It may be that at these low flux values,
Poynting-Robertson is increasing the accretion rate at radii $\la
10$~$r_g$, thereby lowering the surface density and suppressing
reflection features.

\section{Local Changes in Disk Structure}
\label{sect:heating}
The large X-ray flux released during a superburst or Type~I X-ray burst
will significantly heat the surface of the surrounding accretion
disk. In this section we consider how the disk may respond to this
heating\footnote{This was discussed briefly by \citet{bs04}, but is
  presented more fully here. Values derived here supersede the earlier
ones.}.

For a Thomson-thin plasma dominated by photoionization, the maximum
equilibrium temperature obtainable (although free-free
  cooling will somewhat reduce the maximum temperature) is the Compton
temperature $kT_{\mathrm{C}}$,
\begin{equation}
4kT_{\mathrm{C}} \int_{E_1}^{E_2} u_E dE = \int_{E_1}^{E_2} u_E E dE,
\label{eq:comptt}
\end{equation}
where Compton heating is balanced by Compton cooling. In the above
expression $u_E$ is the spectral energy density of the radiation, and
$E$ is photon energy. For the case of a blackbody spectrum produced by
a burst on a neutron star, $u_E \propto E^3/(e^{E/kT}-1)$ and
$kT_{\mathrm{C}} \approx kT$. In response to this increased
temperature, the surface layers of the disk will have to adjust in
order to maintain hydrostatic balance. The timescale of this
readjustment is of the order of the dynamical time $t_z \sim t_{\phi}
\sim 4\times 10^{-6} (M/M_{\odot}) (R/r_g)^{3/2}$~s \citep*{fkr02},
which is essentially instantaneous at distances close to the neutron
star. The new hydrostatic configuration will evolve to a
larger disk scale height $H \propto c_{s} r^{3/2} \propto T^{1/2}
r^{3/2}$, implying that the disk thickness will closely track the hardness of
the burst spectrum. This is interesting because the \fouru\ X-ray reflection
results show a correlation with the observed $kT$ of the incident
power-law. Using standard theory \citep{ss73}, the maximum accretion disk
temperature for \fouru\ is $0.51\ (d/6.4\
\mathrm{kpc})^{-1}$~keV. The temperature of the blackbody spectrum
peaked at $\sim 2.9$~keV \citep{sb02}, so that at this point the
scale-height of the disk atmosphere may have increased by a factor of
at least 2.

\begin{figure}
\centerline{
\includegraphics[angle=-90,width=0.9\textwidth]{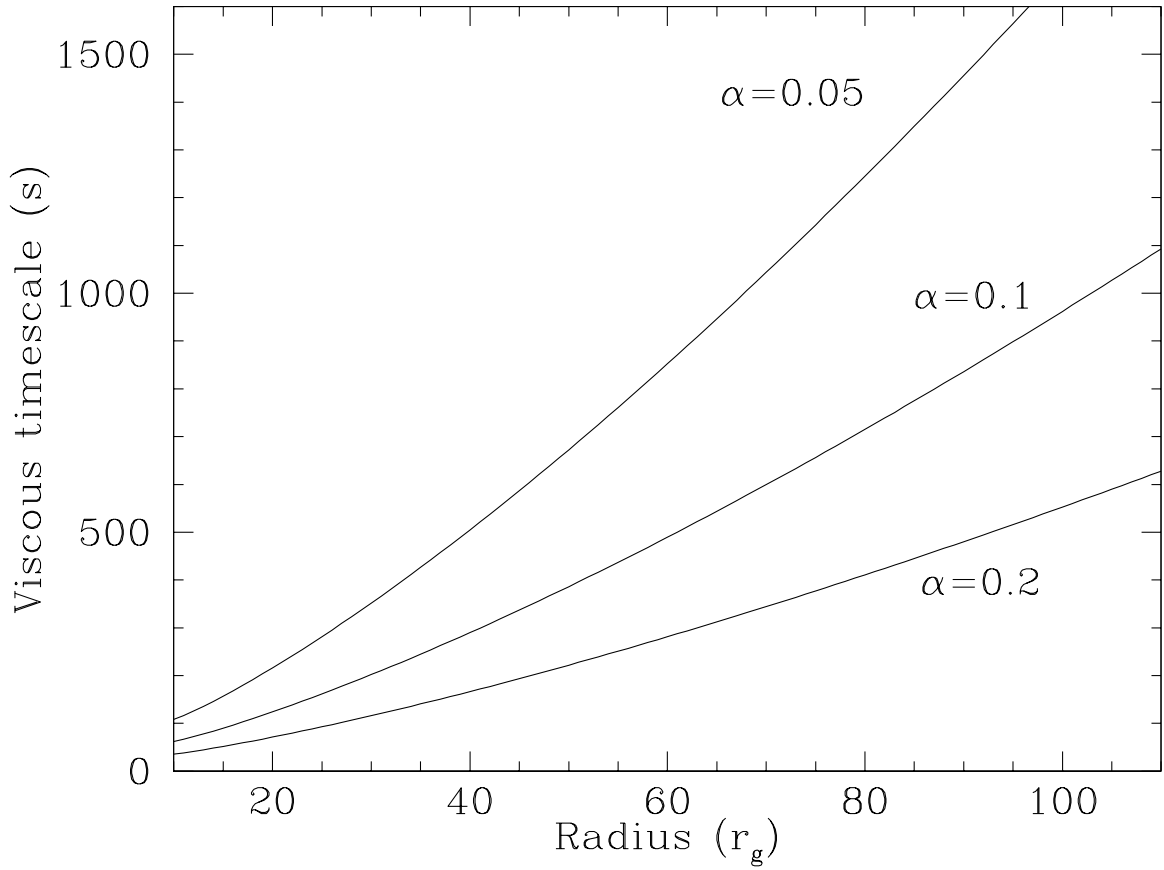}
}
\caption{Viscous timescale for a gas pressure supported disk with
  electron scattering opacity as a function of disk radius. The three
  curves denote different values of the Shakura-Sunyaev viscosity
  parameter. A distance to \fouru\ of 6.4~kpc was assumed, although
  use of the other two distances do not significantly change the
  results. The curves were calculated with Eq.~\ref{eq:tvisc2}, and
  thus assume an unaltered scale height $H$.}
\label{fig:tvisc}
\end{figure}

Significant changes to the surface density due to internal viscosity
mechanisms occur on the viscous timescale, $t_{\mathrm{visc}}$, which
is typically much longer than the dynamical time \citep{fkr02}. For
the gas pressure supported, electron scattering dominated disk
described in \S~\ref{subsub:estimates},
\begin{equation}
t_{\mathrm{visc}} \sim 4.9\times 10^{24} (0.2 (1+X))^{-1/5} (\mu
m_\mathrm{p})^{4/5} \alpha^{-4/5} \left ( {M \over \mathrm{M_{\odot}}} \right
)^{8/5} \dot M^{-2/5} \left ( {R \over r_g} \right )^{7/5}
J(R)^{-2/5}\ \mathrm{s},
\label{eq:tvisc}
\end{equation}
where the symbols are the same as in
eq.~\ref{eq:surfdens}. Substituting values appropriate for \fouru\ (see
\S~\ref{sect:overview} \& \ref{subsub:estimates}), we obtain
\begin{equation}
t_{\mathrm{visc}} \sim 0.22 \alpha^{-4/5} \left ( {M \over 1.4\
\mathrm{M_{\odot}}} \right )^{8/5} \left ( {\dot M \over 1.1\times
10^{17}\ \mathrm{g}\ \mathrm{s}^{-1}} \right )^{-2/5} \left ( {d \over
\mathrm{6.4\ kpc}} \right )^{-4/5} \left ( {R \over r_g} \right
)^{7/5} J(R)^{-2/5}\ \mathrm{s}.
\label{eq:tvisc2}
\end{equation}
Figure~\ref{fig:tvisc} plots $t_{\mathrm{visc}}$ as a function of
radius, assuming three different values of the viscosity parameter
$\alpha$ and a distance of 6.4~kpc (the other distances do not greatly
affect the results). The viscous timescales predicted for $\alpha \sim
0.1$--$0.2$ are in the appropriate range to potentially explain the
evolution of the X-ray reflection features between 20 and
100~$r_g$. The effective $\alpha$ can of course vary from region to
region in the disk and from time to time \citep{hk01,arc01,haw01},
so this should be considered an average value.

The viscous time can also be written as $t_{\mathrm{visc}} \sim
\alpha^{-1} (H/R)^{-2} t_{\phi}$, which illustrates an inverse
dependence on the disk scale height. The estimates of
$t_{\mathrm{visc}}$ from equation~\ref{eq:tvisc2} assumed an unaltered
scale height. As noted above, X-ray heating could increase $H$ by
$\sim 2$, thereby reducing the viscous time by $\sim 4$. If 
$\alpha$ is now assumed to be $\sim 0.025$--$0.05$ then this would
maintain roughly the same $t_{\mathrm{visc}}$ as those plotted in
Fig.~\ref{fig:tvisc}. As $t_{\mathrm{visc}}$ is changing as the gas is
being heated, it may be difficult to obtain an accurate estimate of
$\alpha$.

Since equation~\ref{eq:tvisc2} predicts approximately the correct timescale
for the observed evolution in \fouru, it strongly suggests that the
superburst caused a significant change in the disk structure which is
then propagated on a viscous time. X-ray heating will cause the disk scale
height to increase, but this will occur locally very quickly. We
hypothesize that the sudden, intense and relatively long-lasting X-ray
illumination from a superburst can affect the angular momentum transport
mechanism of the disk.

\section{Discussion}
\label{sect:discuss}
\subsection{The Response of the Accretion Disk around \fouru}
\label{sub:response}
This paper has discussed mechanisms that could influence the structure
or dynamics of an accretion disk around a neutron star undergoing a
Type I X-ray burst or a superburst. The motivation was to determine an
explanation for the apparent evolution of the accretion disk around
\fouru\ during a superburst \citep{bs04}. Here, we will look at the
different possibilities identified above and determine their
effectiveness in explaining the observations.

A radiative driven wind off the accretion disk is an effective means
to remove mass from the inner regions of the disk. However, this
scenario faces several difficulties when confronted with the
data. First, the outflow can only be driven by super-Eddington
luminosities. As mentioned in \S~\ref{sect:overview}, X-ray bursts are
not expected to greatly exceed $L_{\mathrm{Edd}}$, and so the
superburst would have only reached that luminosity for a very short time
at the initial stage of the explosion. \citet{bs04} found that the
X-ray reflecting region drastically changed $\sim 500$~s into the
burst, when the luminosity would have been sub-Eddington. Second, the
mass outflow rates predicted by the numerical wind calculations are
too small to remove the inner disk (assuming it is a Shukura-Sunyaev
disk) unless it subtends a significant fraction of the sky as seen
from the X-ray source. Furthermore, the accretion rate through the
disk is of the same order or even greater than the predicted outflow
rates, and thus increases the difficulty in removing the inner disk. 

On the other hand, it may not be necessary to completely excavate the
inner disk in order to remove the reflection features. All that is
required is to replace the optically-thick, efficiently reflecting
disk with something that is a poor reflector. If a wind is blown off
the disk, then it may obscure the surface from the X-rays and thereby
destroy the reflection features. The longer observed timescale may be
related to this more inefficient evaporation of the disk. More
detailed simulations are required to test this idea.

In \S~\ref{sect:inflow} we found that Poynting-Robertson drag is an
efficient mechanism to remove angular momentum from the gas. However,
the predicted infall timescale is so rapid that the superburst would
drain the entire accretion disk onto the star in only a few
seconds. Clearly, the simple analytic estimate employed in
\S~\ref{sect:inflow} is not applicable to the case of optically thick
accretion disks. One important issue with the analysis is the single
particle assumption in equation~\ref{eq:t2}. If the illuminated gas on
the disk is significantly coupled by the viscous stresses (e.g.,
magnetically) to other areas in the disk, then the mass and
cross-section of this larger, more massive parcel of gas would replace
the single particle in eq.~\ref{eq:t2}. The details of this coupling
would obviously depend on the viscous physics within the accretion
disk, but could produce a mechanism to reduce the angular momentum
loss rate in this material. A full radiation hydrodynamics simulation
which included radiative transfer would be needed to completely
elucidate the effects of radiation drag on the disk material. It seems
likely, therefore, that radiation drag will only be important for the
very innermost regions of the accretion flow (as concentrated on by
\citealt{ml96}). The high luminosity of the burst will very quickly
reduce the angular momentum from this material and increase the local
accretion rate. We would expect the decay timescale to increase
dramatically beyond this small range of radii. This is a potential
explanation for why the inner radius of the reflecting region never
falls below $\sim 10$~$r_g$ (see also \S~\ref{sub:magnetic} below).

The final mechanism considered in the paper involved neither outflow
nor inflow (although it seems likely that both are occurring at some
level), but rather a change in disk structure or surface density due
to X-ray heating. This scenario is the easiest to explain the
correlations seen in the X-ray data, as it predicts a strong
dependence on the blackbody temperature (rather than the flux), and
the viscous timescale is of the right order (with $\alpha \sim
0.03$--$0.2$) to account for the trends seen in the reflections
fits. However, it is the hardest to justify theoretically because the
X-ray heating would change the local disk structure on a dynamical
time, much faster than the viscous time. We suggest therefore that the
intense and long lasting heating does have a large impact on the disk
transport mechanisms which will change on the appropriate timescale. If the
surface density drops globally over the inner accretion disk, due to
the material being 'puffed up' by X-ray heating, then it could easily become
an inefficient reflector. Unfortunately, to truly determine if this
mechanism is the correct explanation, and to determine its importance
relative to the inflow and outflow scenarios described above,
a full numerical simulation which encompasses all the relevant physics
will need to be performed. This is planned for future work.

Of course, there is the possibility that the spectral features
identified by \citet{sb02} are not due to reflection at
all. \citet{st02} argue that since the disk will be heated very
quickly to the blackbody temperature by the burst, all the heavy metals
will be entirely ionized and unable to produce reflection
features. These authors propose that the \fe\ line and edge are formed
in an outflow from the stellar surface \citep{ts04}. While it is
possible that spectral features may form in an outflow (although it
would be unusual to find a large column of heavy metals close to the
surface of a neutron star; \citealt{bsw92}; \citealt{bcp03}) it is not
clear that the disk will be entirely ionized by the
superburst. Numerical simulations of accretion disks subject to the
magnetorotational instability consistently find that the gas is
inhomogeneous and clumpy due to turbulence \citep[e.g.,][]{hk01}. These density
inhomogeneities may enhance the line emission in the X-ray
reflection spectrum \citep{btb04}.

\begin{figure}
\centerline{
\includegraphics[angle=-90,width=0.9\textwidth]{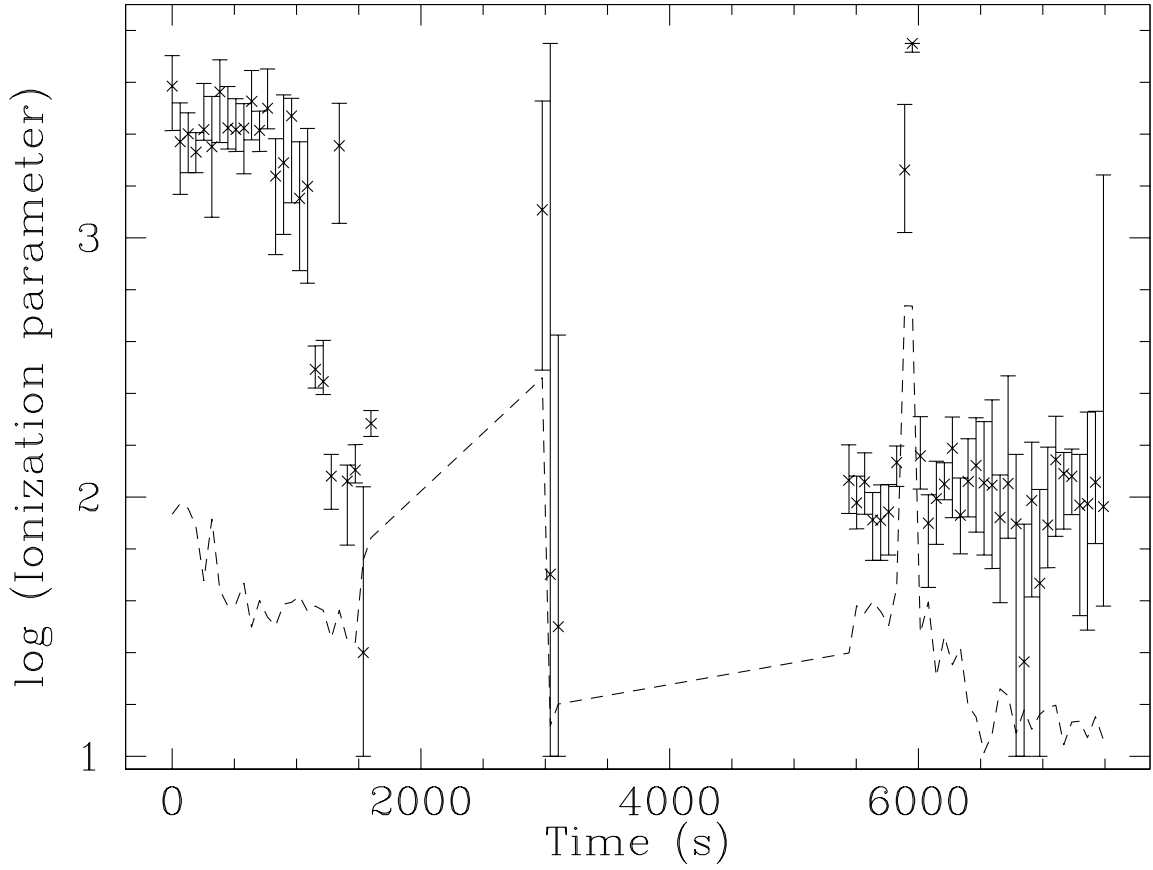}
}
\caption{The dashed line plots the predicted value of the disk
  ionization parameter from equation~\ref{eq:xi2} during the \fouru\
  superburst. The points are the best fit values from the reflection
  modeling of \citet{bs04}. A distance of 6.4~kpc and $\alpha=0.1$
  was assumed for the theoretical estimate. In order to bring the
  prediction in line with the observations, the surface density must
  be decreased by about a factor of 40 at the beginning of the burst,
  but this decreases to $\sim 8$ at later times.}
\label{fig:ionispar}
\end{figure}

The spectral fitting by \citet{bs04} measured the ionization parameter
$\xi=4\pi F_{\mathrm{X}}/n_{\mathrm{He}}$ of the reflecting region on
the disk. Shakura-Sunyaev disk theory can be used to
estimate $\xi$ and determine if the disk is expected to be completely
overionized. Writing $n_{\mathrm{He}}=\Sigma/m_{\mathrm{He}}H$ and
$F_{\mathrm{X}}=L/4\pi R^2$, we obtain
\begin{equation}
\xi \approx {L m_{\mathrm{He}} H \over R^2 \Sigma}.
\label{eq:xi1}
\end{equation}
The scale-height $H$ for a gas-pressure dominated disk with electron
scattering opacity is
\begin{equation}
H = 1.4\times 10^{-10} (\mu m_{\mathrm{p}})^{-2/5} (0.2 (1+X))^{1/10}
\alpha^{-1/10} \left ({M \over \mathrm{M_{\odot}}} \right )^{7/10}
\left ( {R \over r_g} \right )^{21/20} \dot M^{1/5}
J(R)^{1/5}\ \mathrm{cm}.
\label{eq:height}
\end{equation}
Using this expression, equation~\ref{eq:surfdens}, $X \approx 0$ and
$\mu \approx 1.3$, we now obtain
\begin{footnotesize}
\begin{equation}
\xi \approx 1500 \alpha^{7/10} \left ( {L \over
  L_{\mathrm{Edd}}} \right ) \left( {M \over 1.4\ \mathrm{M_{\odot}}}
  \right )^{-9/10} \left ( {R \over r_g} \right )^{-7/20}
  \left ({\dot M \over 1.1\times 10^{17}\ \mathrm{g}\ \mathrm{s}^{-1}}
  \right )^{-2/5} \left ( {d \over 6.4\ \mathrm{kpc}} \right )^{-4/5} J(R)^{-2/5}\ \mathrm{ergs\ cm\ s^{-1}}.
\label{eq:xi2}
\end{equation}
\end{footnotesize}
Figure~\ref{fig:ionispar} plots the predicted $\xi$ at the
radius measured by the spectral fitting of \citet{bs04}, assuming it
is illuminated by the luminosity plotted in
Fig.~\ref{fig:loverledd}. Following the results of
\S~\ref{sect:heating}, we have also assumed $\alpha =
0.1$. Overplotted on the figure are the measured values of the
ionization parameter as inferred from the spectral fitting.

In contrast to the expectations of \citet{st02}, the disk does not
have to be overionized by the burst. Rather, Figure~\ref{fig:ionispar}
shows that the predicted $\xi$ is generally much smaller than the
measured one, especially near the beginning of the superburst. A
simple way to bring the theoretical prediction of $\xi$ in line with
the measured one is to have a variable surface density which is $\sim
40$ times smaller over the first $\sim 1000$~s, but then becomes only
$\sim 8$ times smaller a couple of hours later, as the burst is
decaying. Of course, increasing the scale height also results in a
larger $\xi$. The exact behavior of how the surface density would
evolve as a function of radius and burst luminosity would have to be
calculated with a numerical simulation. Finally,
Figure~\ref{fig:ionispar} shows that the predicted $\xi$ follows the
same trend as those inferred from the spectral fitting, supporting the
ionized reflection interpretation of the observed spectral features.

There are additional dynamical effects that arise in bursting systems
that could affect the accretion flow. For example, while models of PRE bursts
show that the luminosity from the photosphere will barely exceed the
Eddington limit, they also indicate that winds will be driven from the
surface of the star
\citep*{ehs83,kat83,kat85,qp85,pa86,pp86,jm87}. Thus, while the burst
radiation may not blow away the disk, there may be a mechanical
interaction between the stellar wind and the accretion flow. However,
the outflows generally carry very little kinetic energy compared to
the radiative energy released in the burst, so any mechanical
interaction is unlikely to be significant. Secondly, the X-ray heated
gas on the disk surface will want to move to larger radii due to the
gas pressure gradient as well as radiation pressure. This tendency to
rotate at larger radii will be counteracted by viscous transport within
the hot gas and radiative drag. Therefore, there may be some
interesting dynamics within the disk as a response to X-ray bursts.

\subsection{Disk Truncation and Magnetic Effects}
\label{sub:magnetic}
Modeling of the reflection features in the spectra of the \fouru\
superburst showed that there was little to no reflection within a
radius of $\sim 15$~$r_g$ \citep{bs04}, much larger than the innermost
stable circular orbit at $6$~$r_g$. One explanation for this
apparently truncated accretion flow is locally enhanced accretion due
to Poynting-Robertson drag (\S~\ref{sect:inflow}). Alternatively, this
radius may correspond to the magnetospheric radius, $R_m$, where the
magnetic pressure from the neutron star's magnetic field is equal to
the gas ram pressure in the accretion flow. Within $R_m$ the dynamics of the
accreting gas is dominated by the magnetic field, and therefore the
disk may be easily disrupted into a configuration which is unlikely to
produce reflection features.

The magnetospheric radius is \citep{gl78,gl79}
\begin{equation}
\left ( {R_M \over r_g} \right ) \approx 1.3\times 10^{-9} \left ( {M
  \over \mathrm{M_{\odot}}} \right )^{-8/7} \dot M^{-2/7} \mu_m^{4/7},
\label{eq:rm}
\end{equation}
where $\mu_m$ is the magnetic moment of the stellar field (assuming
dipole geometry). Substituting $M=1.4$~M$_{\odot}$ and $\dot M \approx
1.1\times 10^{17} (d/6.4\ \mathrm{kpc})^2$~g~s$^{-1}$, and taking $r_m
\approx 15$~$r_g$, we find that $\mu_m \approx 2.8\times 10^{26}\
(d/6.4\ \mathrm{kpc})$~G~cm$^{3}$, or $B \sim 3\times 10^8\ (d/6.4\
\mathrm{kpc})$~G. This is a reasonable field strength for a
non-pulsating LMXB system \citep{pc99}. In fact, almost all systems
which exhibit X-ray bursts, and LMXBs in general, do not show X-ray
pulsations \citep{lvt93}, implying much weaker field strengths then
pulsars (which have $B \sim 10^{12}$~G). Thus, it is possible that the
lack of X-ray reflection from within $\sim 15$~$r_g$ may be due to
magnetic truncation.

\subsection{Prospects for Future Observations}
\label{sub:future}
The superburst from \fouru\ occurred when the system was being
observed by \textit{RXTE}. As a result, the sensitive narrow-band
instruments onboard were able to pick up the reflection features and
track their evolution as the burst evolved. So far only one other
system has been as fortunate (4U~1636--53; \citealt{sm02};
\citealt{ku04}).  Superbursts are the ideal events for which to search
for X-ray reflection features from the accretion disk as they last for
a number of hours. Moreover, because of this long timescale, the
analyses presented
in this paper predict that we are more likely to observe changes to
the accretion disk during superbursts than during the much shorter Type~I
X-ray bursts. The major difficulty is the unknown recurrence time of
superbursts which can vary over a number of years depending on the
composition of the accreted material \citep{cum03}. Nevertheless, we suggest
that rapid X-ray followup of superbursts would be an important
secondary science objective for the \textit{Swift} mission. Detecting
reflection features from a variety of systems with different orbital
and spin periods, magnetic field strengths and evolutionary paths
would allow examination of the accretion process over a wide parameter space.

As the overall geometry is the same and the light travel time is much
shorter than the burst duration, X-ray reflection features should be
produced in Type I bursts as well. The difficulty will be detecting
them with sensitive enough instruments to do time averaged spectral
analysis let alone time-resolved studies. However, Type I events are
much more common than superbursts allowing repeatable experiments and
tests of hypotheses. Sensitive spectral analysis of Type~I X-ray
bursts would be an important objective for a future high-throughput
observatory.

\section{Conclusions}
\label{sect:concl}
Type I X-ray bursts or superbursts on neutron stars produce a
significant amount of energy ($\sim 10^{39}$ or $\sim 10^{42}$~ergs,
respectively) in a short amount of time. In this paper, we have
discussed how this energy, in the form of a X-ray flux, would affect
the surrounding accretion flow. The goal was to elucidate the meaning
behind the apparent evolution of the accretion disk uncovered by
\citet{bs04} in their analysis of X-ray reflection features observed
during the \fouru\ superburst.

Three separate physical processes were considered. The first,
radiatively driven winds, was found to be occur, but only for
$L/L_{\mathrm{Edd}} \ga 1$ and, for a thin accretion disc, the
predicted outflow rate was less than the accretion rate, so this
effect may not produce a large change to the accretion
flow. Poynting-Robertson drag, the second process, was in fact
predicted to be too efficient, and would have removed the inner
accretion disk is less than a second. This was the result of using a
too simple description of the problem, but Poynting-Robertson may be
dominant very close to the central star. The final process considered
was a change in surface density due to X-ray heating. This
successfully reproduced the observed timescale for a Shakura-Sunyaev
viscosity parameter of $\alpha\sim 0.03$--$0.2$, but requires the
\emph{ansatz} that the X-ray heating will result in significant
changes to the surface density of the disk that would propagate on a
viscous time. Our conclusion is that, while the X-ray heating may
dominate the explanation of the observations, all three processes will
be ongoing during a superburst, and numerical simulations are required
to fully understand the response of the accretion disk to such large
explosions.

Finally, we emphasize that future rapid follow-up observations of
X-ray bursts (both super and otherwise) will be vital in the progress
of this problem. X-ray reflection features should be common in these
systems and any observed changes are likely to correspond to changes
in the accretion flow. Further examples are needed to test the ideas
presented here as well as provide constraints on accretion theory.

\acknowledgments

The authors thank C. Matzner and T. Strohmayer for useful discussions and
comments, and the anonymous referee for suggestions which improved the
presentation of the paper. This research was supported by the Natural
Sciences and Engineering Research Council of Canada.

\clearpage

\clearpage

\clearpage

\clearpage

\clearpage

\clearpage

\end{document}